\documentclass[12pt]{article}

\usepackage{placeins}
\usepackage{pifont}
\usepackage{graphicx}
\usepackage{booktabs}
\usepackage{multirow}
\usepackage{amsmath}
\usepackage{amssymb}
\usepackage{lmodern}
\usepackage{rotating}
\usepackage{float}
\usepackage[T1]{fontenc}
\usepackage[a4paper,left=2.5cm,right=2.5cm,top=3cm,bottom=3cm]{geometry}
\usepackage[flushleft]{threeparttable}

\usepackage[style=authoryear-comp, backend=bibtex, maxcitenames=2, maxbibnames=99 ,dashed=false]{biblatex}
\bibliography{covid_impf}

\usepackage{hyperref}

\begin{document}

\title{Impact of Record-Linkage Errors in Covid-19 Vaccine-Safety Analyses using German Health-Care Data: A Simulation Study}
\author{Robin Denz$^1$, Katharina Meiszl$^1$, Peter Ihle$^2$,\\ Doris Oberle$^3$, Ursula Drechsel-Bäuerle$^3$, \\ Katrin Scholz$^2$, Ingo Meyer$^2$, Nina Timmesfeld$^1$}
\date{\small $^1$Ruhr-University Bochum, Department of Medical Informatics, Biometry and Epidemiology \\ $^2$University of Cologne, PMV Research Group \\ $^3$Paul-Ehrlich-Institut, Federal Institute for Vaccines and Biomedicines Germany, Department Safety of Biomedicines and Diagnostics}

\maketitle

\begin{abstract}
	With unprecedented speed, 192,248,678 doses of Covid-19 vaccines were administered in Germany by July 11, 2023 to combat the pandemic. Limitations of clinical trials imply that the safety profile of these vaccines is not fully known before marketing. However, routine health-care data can help address these issues. Despite the high proportion of insured people, the analysis of vaccination-related data is challenging in Germany. Generally, the Covid-19 vaccination status and other health-care data are stored in separate databases, without persistent and database-independent person identifiers. Error-prone record-linkage techniques must be used to merge these databases. Our aim was to quantify the impact of record-linkage errors on the power and bias of different analysis methods designed to assess Covid-19 vaccine safety when using German health-care data with a Monte-Carlo simulation study. We used a discrete-time simulation and empirical data to generate realistic data with varying amounts of record-linkage errors. Afterwards, we analysed this data using a Cox model and the self-controlled case series (SCCS) method. Realistic proportions of random linkage errors only had little effect on the power of either method. The SCCS method produced unbiased results even with a high percentage of linkage errors, while the Cox model underestimated the true effect.
\end{abstract}

\section{Introduction}

The rapid development and deployment of Covid-19 vaccines has been pivotal in the global effort to fight the Covid-19 pandemic. In Germany alone, 192,248,678 doses of different vaccines have been used as of July 11, 2023 \parencite{RKI2023}. Because of this unprecedented magnitude and the continued usage of these vaccines, it is particularly important that their safety is closely studied. Although randomised trials have been conducted to investigate the efficacy and safety of the vaccines \parencite{Polack2020, Sadoff2021, Falsey2021, Baden2021}, these trials are not sufficient to obtain a full safety profile of the vaccines before marketing, due to their strict inclusion criteria and sample sizes \parencite{Hartzema1987}. Instead, routine health-care data may be used to identify possible adverse side-effects in a more realistic setting \parencite{Schuemie2022, Eitelhuber2022}.
\par\medskip
Despite the large number of individuals enrolled in health insurances, using health-care data for vaccine-safety analysis is not a trivial task. Aside from general problems of data protection, data quality and representativeness \parencite{Dash2019, Awrahman2019}, the main problem when analysing Covid-19 vaccination-related questions in Germany is that an individual's Covid-19 vaccination status is recorded independently from the person's other health-care data. The main data source for Covid-19 vaccinations is the Digitales Impfquoten-Monitoring database \parencite{RKI2023}, which is currently not linked to the data of any health-care provider. Since there is no persistent, unique and database-independent person identifier in Germany, error-prone probabilistic record-linkage methods have to be used to merge these databases.
\par\medskip
Record-linkage methods usually rely on pseudo person-identifying features such as names, addresses and birth-dates to identify individuals across multiple databases \parencite{Christen2020}. Due to the possibility of data entry errors such as misspelled names or dates and because of the possibility that some of these features change over time, linkage errors cannot be completely prevented in practice, even with state of the art methods \parencite{Bohensky2010, Harron2014}. A detailed description of record-linkage in general and possible sources of errors therein can be found in \textcite{Christen2020}. In this article however, we are not concerned with the reasons for the errors per-se. Instead we are only interested in what \emph{kind} of linkage problems could be present in the resulting merged data and, more importantly, what the impact of these errors is when analysing vaccine-safety.
\par\medskip
Linkage errors are often ignored in practice, which may lead to biased estimates and/or a loss of statistical power \parencite{Shaw2022, Bohensky2010, Doidge2019}. The actual extent of these problems depending on the type and probability of linkage errors is largely unknown. Although this topic has been studied in the past, most of the existing studies focus on representativeness and misclassification. To the best of our knowledge, there is no study dealing explicitly with the impact of record-linkage errors in vaccine-safety analyses. Additionally, most of the present research relies exclusively on real data \parencite{Shaw2022, Parrish2017, Moore2016}, with a few notable examples \parencite{Harron2014, Baldi2010}. Real data can be used effectively to showcase empirical problems, but it usually does not allow the assessment of bias, because the true data generation process is unknown in most cases. Additionally, it does not allow the study of multiple different hypothetical scenarios.
\par\medskip
We use a comprehensive Monte-Carlo simulation to study the impact of record-linkage errors on the bias and statistical power when employing two different analysis strategies. In particular, we utilize a \emph{Cox proportional hazards regression model} \parencite{Cox1972} with time-varying covariates and the \emph{self-controlled case series method} \parencite{Farrington1995} to assess vaccine-safety in the simulated data in settings with different kinds and amounts of record-linkage errors. We then assess the bias and statistical power in each scenario. The main disadvantage of this approach is, however, that the generated data may be too simple or generally unrealistic, which may lead to un-generalizable results. We try to mitigate this problem by using a \emph{discrete-time simulation} approach \parencite{Tang2020} where empirical data is used directly to tune the relevant parameters of the data-generation process. In particular, we rely on the fairly well-documented effect of mRNA vaccines on the occurrence of acute myocarditis \parencite{HromicJahjefendic2023}.

\section{Methods}

Because we are only interested in the effects of the record-linkage errors and not on the process by which they occur, we only generate a single dataset per simulation replication. This dataset resembles the outcome of a perfect linkage process at first, containing all relevant information to conduct the study of interest such as vaccination status and event status. The dataset is then augmented to include the different kinds of record-linkage errors described below. Afterwards, the different statistical analysis methods are carried out on this manipulated dataset and the results are recorded. In the following section we will give a detailed description on each of the constituting parts of this simulation.

\subsection{Target Estimand \& Statistical Analysis Methods} \label{chap::target_estimand}

We assume that there is a single relative risk of developing an adverse event of interest in the $d_{risk}$ days at risk after receiving an mRNA based Covid-19 vaccination compared to the time not at risk, denoted by $RR_{vacc}$, which is fixed over time and applies to every person equally. This $RR_{vacc}$ is the quantity we aim to estimate. Let $P_0(t)$ be the time-dependent baseline probability of developing the adverse event of interest and let $T_{vacc}$ be the time at which a person receives a Covid-19 vaccination. The probability of developing the event at $t$ is then given by:

\begin{equation}
	P(t) = 
	\begin{cases}
		P_0(t) \cdot RR_{vacc}, & \text{if } t \in  \left[T_{vacc}, T_{vacc} + d_{risk}\right] \\
		P_0(t), & \text{otherwise}
	\end{cases}
\end{equation}

This equation could be adapted to allow different relative risks per number of vaccines received per person, by adding more specific conditions. We assume a single $RR_{vacc}$ here, but model the first and second vaccination separately in the models because this is the approach usually taken in real data analysis. Multiple different statistical methods have been proposed to identify $RR_{vacc}$ from observational data, such as the \emph{self-controlled case series} (SCCS) method \parencite{Farrington1995}, the \emph{Cox proportional hazards regression} model \parencite{Cox1972}, \emph{sequence symmetry analysis} \parencite{Lai2017} and the \emph{case-crossover design} \parencite{Zhang2016}, among others. In this article we will focus only on the first two methods mentioned, because they are often used in pharmacovigilance. A small review and simulation study comparing some of these methods can be found in \textcite{Glanz2006} and \textcite{Takeuchi2018}. 
\par\medskip
The Cox model is given by:

\begin{equation}
	h(t) = h_0(t)\exp(\beta X(t)),
\end{equation}

where $\beta$ is a vector of coefficients, $X(t)$ is a matrix of (possibly) time-dependent covariates and $h_0(t)$ is the baseline-hazard function, which is usually left unspecified \parencite{Cox1972}. It is possible to include time-varying covariates, such as the vaccination status, in this model by including multiple time-periods per person as described in \textcite{Zhang2018a}. To estimate the constant relative risk in the post-vaccination risk period, a dichotomous covariate is introduced which is equal to one in the $d_{risk}$ days after the vaccination and zero otherwise. This covariate is then simply included as an additional independent variable. We use the time until the first event of interest occurs as outcome throughout the article. Although the Cox model does not estimate relative risks directly, the estimated hazard ratios from the model can be used to approximate the relative risks if the event is rare \parencite{VanderWeele2020}.
\par
\begin{figure}[!htb]
	\centering
	\includegraphics[width=1.04\linewidth]{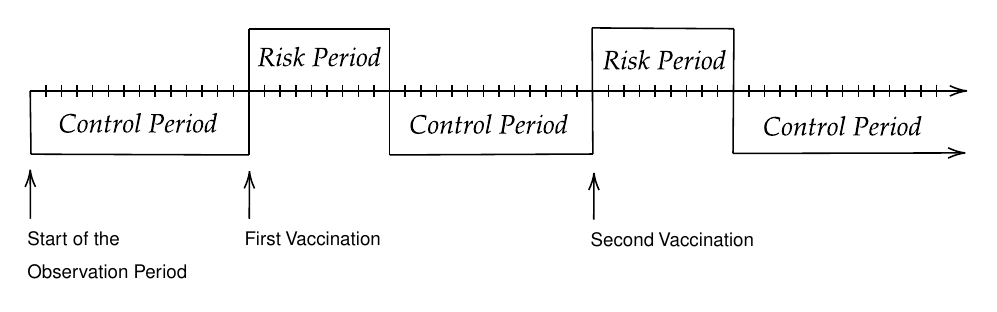}
	\caption{Schematic representation of the different time-periods for a fictional individual in the SCCS design.}
	\label{fig::sccs}
\end{figure}

\medskip
The SCCS method was originally created specifically to study the effect of vaccinations on the occurrence of adverse events, but has been used in various other types of studies as well \parencite{Farrington1995, Whitaker2006}. In an SCCS design only individuals who experienced the adverse event of interest are included. For each person the time in the risk period after the vaccination is then compared with the rest of the persons observation time. Because each person acts as their own control, all time-invariant person-specific covariates are implicitly controlled for automatically. Figure~\ref{fig::sccs} shows how the person-specific observation time is divided for a fictional individual.

\par\medskip
The Cox model utilizes information about all individuals, regardless of vaccination and event status, while the SCCS method relies solely on information about individuals which have experienced the event of interest at some point in time. The former is therefore sometimes called a \emph{cohort design}, while the latter is often called a \emph{case-only design} \parencite{Takeuchi2018}. The main difference between the two approaches is that in the Cox model vaccinated individuals are compared to unvaccinated people to obtain the estimate, while the SCCS method compares different time-periods within each person \parencite{Whitaker2006}. The SCCS method has the advantages that time-invariant confounders (measured and unmeasured) are automatically controlled for. It is also less computationally demanding, because the dataset used is much smaller than in the cohort design. Because it uses the time before and after the event as control periods, it may, however, be more susceptible to immortal time bias \parencite{Maclure2012}. Consequently, record-linkage errors may influence the estimates produced by these two methods in different ways.

\FloatBarrier

\subsection{Data Generation}

To generate the data required for the subsequent Monte-Carlo simulation consisting of $n_{sim}$ individuals with information about $n_{days}$ simulated days, we used a \emph{discrete-time simulation} approach \parencite{Tang2020}. In a discrete-time simulation an initial population of $n_{sim}$ individuals with some characteristics at baseline is generated. Afterwards the simulation time $t$ is increased by one at each of $n_{days}$ steps. With each time increase the state of the individuals of the population is updated according to a arbitrarily complex user-specified mechanism. All events that occur in one step are recorded. The data-generation is finished after $n_{days}$ steps or when a pre-specified condition is met \parencite{Tang2020}. This method is sometimes also called \emph{dynamic microsimulation} \parencite{Spooner2021} and is closely related to \emph{discrete-event simulation} \parencite{Banks2014}.
\par\medskip
Here, the process starts by generating $n_{sim}$ individuals. Each of these individuals is not vaccinated and has not experienced the event of interest in the past. The simulation time $t$ is then increased by one unit, which corresponds to a time increase of one day. The probability of being vaccinated for the first time is then determined for each individual based on $t$ and used in Bernoulli trials to determine the vaccination status. To make the data as realistic as possible, we relied on empirical estimates of the daily vaccination rate in Germany as reported by the Robert Koch Institute \parencite{RKI2023}. If a person is vaccinated for the first time, the type of vaccine is drawn using a weighted random sample, where the weights correspond to the observed proportions of the possible vaccine types. Those proportions are again used as reported by the Robert Koch Institute \parencite{RKI2023}.
\par
\begin{figure}[!htb]
	\centering
	\includegraphics[width=0.9\linewidth]{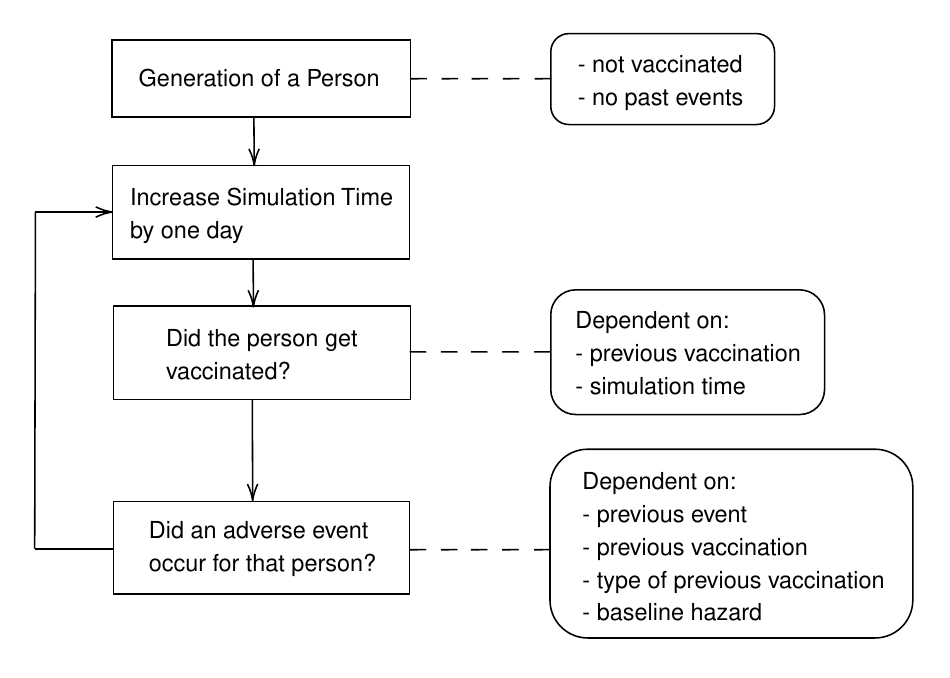}
	\caption{ Flow-diagram of the data-generation process.}
	\label{fig::flow_sim}
\end{figure}

\medskip
If the person has already been vaccinated once, the individual will be vaccinated again at $t + 42$ if the person received either the BionTech or Moderna vaccine and at $t + 84$ if the person received the AstraZeneca vaccine. Individuals who have received the vaccine from Janssen will never receive a second vaccination. This resembles perfect adherence to the official German STIKO guidelines at the beginning of the vaccine campaign \parencite{RKI2023}. No further vaccines are simulated. Once the vaccination status at $t$ has been generated, the probability of experiencing the event of interest is determined. If the individual has been vaccinated using either the BionTech or Moderna vaccine at $t$ or in the previous $d_{risk}$ days, the baseline probability of the event is multiplied by the constant factor $RR_{vacc}$. This constant corresponds to the target estimand as defined in~\ref{chap::target_estimand}. Whether an event occurs or not is again determined through Bernoulli trials. If an event occurs it may occur again for the same individual at a later point in time after some immunity duration $d_{immune}$ is over. Specific values for all simulation parameters are given below. A flow-chart of the data-generation process is shown in figure~\ref{fig::flow_sim}.

\FloatBarrier

\subsection{Possible Linkage Errors}

In this particular instance, there are two main data sources. The first contains data on Covid-19 vaccinations performed in Germany since the 27th of December 2020. Secondly, there are multiple health-care databases created by different German health-care providers, which we will refer to as health-care data. Neither of these database are exhaustive, meaning that the vaccination database does not contain information about every single vaccination dose that was actually administered and that the combined health-care database does not contain information about every German citizen. The goal is therefore not to create a dataset with information about the whole population of Germany. Instead, we merely aim to merge information about the received vaccinations with available health-care data. After linking both data sources, the resulting database would ideally correspond to one entry per person enrolled in German health-care including the complete health-care data per person plus the persons' correct vaccination status.
\par\medskip
Despite the large amount of possible sources of errors in the linkage process \parencite{Christen2020, Boyd2017}, it is possible to categorize the result of these errors into two categories in the present case: \emph{missing matches} and \emph{false matches}. Missing matches occur when a person was vaccinated, but the data about this vaccination could not be linked to the persons health-care data. This may occur when the vaccination was not recorded at all or when some of the identifying information, such as the name or the birth date of the person, is incorrectly recorded in one of the two data sources \parencite{Christen2020}. In the resulting database, this person will incorrectly show up as not vaccinated. False matches on the other hand may occur if the identifying information of the person is not unique. For example, suppose there are two individuals named John Doe, both born on the first of January 1960 in the same city. If only one of these two individuals is included in the health-care data, the vaccination status will be linked partially incorrectly because it is not known that there is another person with the same identifying information.
\par\medskip
In the following simulation study we simulate missing matches by randomly choosing vaccinated individuals from the generated dataset containing all information and setting their vaccination status to ``not vaccinated''. False matches are simulated similarly by randomly choosing a number of individuals and exchanging their vaccination information. By choosing the individuals randomly, we assume that the occurrence of these record-linkage errors is independent of all person-specific attributed, such as age or sex. This is a strong assumption, especially since some studies have identified systematic differences between individuals that could be linked and those that could not be linked in empirical studies \parencite{Boyd2017, Miller2017, Randall2018}. We nevertheless choose to make this assumption to investigate whether record-linkage errors have an impact even without the presence of selection bias.

\subsection{Simulation Parameters \& Scenarios}

We use the adverse effect of mRNA based vaccines on the occurrence of acute myocarditis as a guiding example to determine the relevant simulation parameters. Multiple studies have shown evidence that both the Pfizer-BionTech \parencite{Polack2020} and Moderna \parencite{Baden2021} vaccines raise the probability of developing acute myocarditis, especially in young males, although this evidence is not entirely conclusive \parencite{Khan2022, HromicJahjefendic2023}. All simulation parameters are chosen so that the simulated data is similar to this specific example. Through this approach, we try to generate realistic data which may make the results of this study more generalizable.
\par\medskip
Accordingly, we set the baseline probability of developing acute myocarditis in one year to 0.00016, which corresponds to estimates from the Global Burden of Disease Study 2013 \parencite{GBDSC2015}. The duration of the risk period after receiving either the Pfizer-BionTech vaccine or the Moderna vaccine was set to 21 days and the corresponding relative risk for the development of acute myocarditis was set to 3.24, as estimated by \textcite{Barda2021} and \textcite{Shimabukuro2021}. Additionally, we set the duration in which a person is immune after suffering from acute myocarditis once to 42 days, as determined by a discussion with experts. The simulation is run for a total of 550 days, where the vaccination campaign starts after 365 days. The population size was set to 770,000 individuals, which is approximately the number of people needed to achieve a statistical power of 80\% for detecting a statistically significant difference with $\alpha = 0.05$, as calculated using the method of \textcite{Musonda2006} with the values mentioned above and assuming no linkage errors are present. The whole simulation is repeated 2000 times.
\par\medskip
We were unable to obtain empirical estimates of the probability of the two different linkage errors discussed here. However, we hypothesize that \emph{false matches} are very rare in practice, because they can only occur if at least two persons with identical identifying information are included in the database. The probability of this kind of linkage error is therefore set to a constant 50 cases per one Million people throughout the simulation. Additionally, we hypothesize that \emph{missing matches} are much more likely and systematically vary this parameter in the simulation since no empirical estimate is available. Table~\ref{tab::sim_params} gives a brief overview of all simulation parameters used. Additionally, in accordance with recommendations for discrete-event simulations \parencite{Monks2019}, we give a detailed overview of all assumptions made in the simulation in the appendix.

\begin{table}[!htb]
	\centering
	\label{tab::sim_params}
	\begin{tabular}{lll}
		\toprule
		Parameter & Value & Description \\
		\midrule
		$n_{sim}$ & $770,000$ & Population size \\
		$n_{repeats}$ & $2000$ & Number of simulation replications \\
		$n_{days}$ & $550$ & Number of unique time-steps (days) in the simulation \\
		$d_{risk}$ & 21 & Risk duration after vaccination in days \\
		$d_{immune}$ & 42 & Duration after an event where the event cannot occur again \\
		$RR_{vacc}$ & 3.24 & Relative risk of vaccination \\
		$P_{myocarditis}$ & 0.00016 & Probability of developing an acute myocarditis in a year \\
		$P_{vacc}(t)$ & RKI data & Time-dependent probability of receiving the first vaccination \\
		$P_{vacc\_type}$ & RKI data & Probability distribution determining the type of received \\
		& & vaccine \\
		$P_{False Match}$ & 0.00005 & Proportion of false matches \\
		$P_{Missing Match}$ & Varied & Proportion of missing matches \\
		\bottomrule
	\end{tabular}
	\caption{An overview of the parameters used in the simulation study.}
\end{table}

\FloatBarrier

\subsection{Performance Criteria}

To judge the performance of each considered method we rely on the bias, the mean-squared error (MSE) and the statistical power. The bias is defined as:

\begin{equation}
	Bias = E\left(RR_{vacc} - \hat{RR}_{vacc}\right),
\end{equation}

and estimated by taking the arithmetic mean of the difference between $RR_{vacc}$ and $\hat{RR}_{vacc}$ over all simulation replications. An unbiased estimator will have a theoretical $Bias$ of 0, with the estimated bias tending towards 0 with an increasing number of simulation replications. Similar to the $Bias$, the $MSE$ is defined as:

\begin{equation}
	MSE = E\left(\left(RR_{vacc} - \hat{RR}_{vacc}\right)^2\right),
\end{equation}

and is estimated again by averaging over all simulation replications. Contrary to the $Bias$ the $MSE$ will generally not be equal to 0 for unbiased estimators. Instead it gives an indication on how close the estimated relative risks are to the true relative risk, with extreme outliers being ``punished'' more harshly due to the quadratic term. In general, a more precise estimator will have a smaller mean-squared error.
\par\medskip
In addition to these two metrics, we also use the statistical power to judge the performance of the methods. The statistical power is defined as the probability of correctly rejecting a false null-hypothesis given a pre-specified false-rejection probability $\alpha$. In our case, the null-hypothesis is that $RR_{vacc}$ is equal to 1 and we use $\alpha = 0.05$ throughout the article. We estimate the statistical power by calculating the p-value for the corresponding hypothesis test in each simulation run and recording the proportion of p-values $< 0.05$ over all simulation replications. All estimates are reported with associated Monte-Carlo errors, calculated as the standard error of the estimates over all replications, and associated 95\% confidence intervals. Additionally, we also graphically display the distribution of the estimated relative risks in each of the considered scenarios in the appendix.

\subsection{Software}

The entire simulation study was performed using the \textbf{R} programming language (Version 4.0.2). We used code that was subsequently further developed into the \texttt{simDAG} R-package \parencite{Denz2023} to generate the required data. All code is available in the online appendix of this article.

\section{Results}

\begin{figure}[!htb]
	\centering
	\includegraphics[width=1\linewidth]{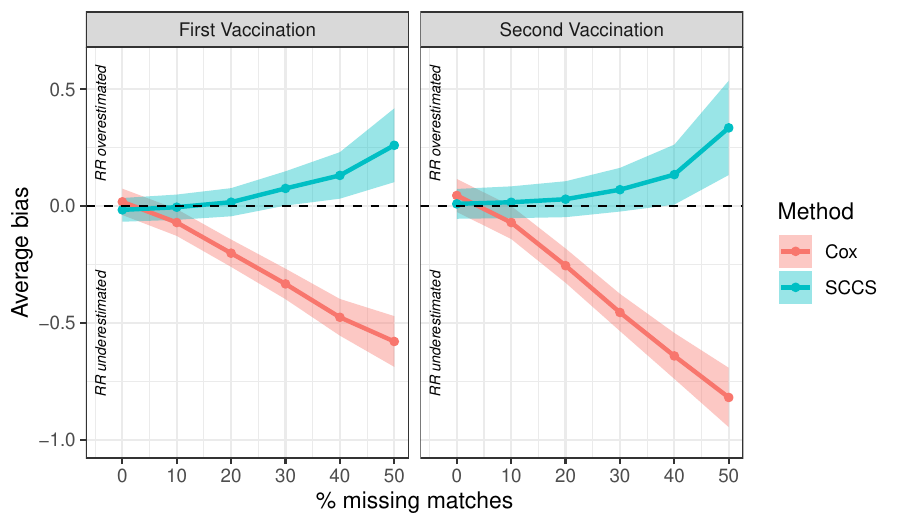}
	\caption{Average Bias of the two evaluation methods in detecting the effect of mRNA vaccination on the occurrence of myocarditis for different proportions of missing matches with 95\% confidence intervals.}
	\label{fig::myokarditis_bias}
\end{figure}

Figure~\ref{fig::myokarditis_bias} displays the average bias of the estimated relative risks for the first and the second vaccination for both the SCCS method and the Cox model with different percentages of missing matches. With no missing matches, both methods produce essentially unbiased results. As the percentage of missing matches increases to ten percent, the bias is still very close to zero when using the SCCS method, but begins to become negative for the Cox model, meaning that the Cox model underestimates the true effect of the vaccination. With 20\% missing matches the bias is still almost non-existent for the SCCS method, but grows sharply for the Cox model. At higher percentages of missing matches, SCCS method starts to overestimate the relative risk. This trend is observed for both the first and the second vaccination.
\par\medskip

\begin{figure}[!htb]
	\centering
	\includegraphics[width=1\linewidth]{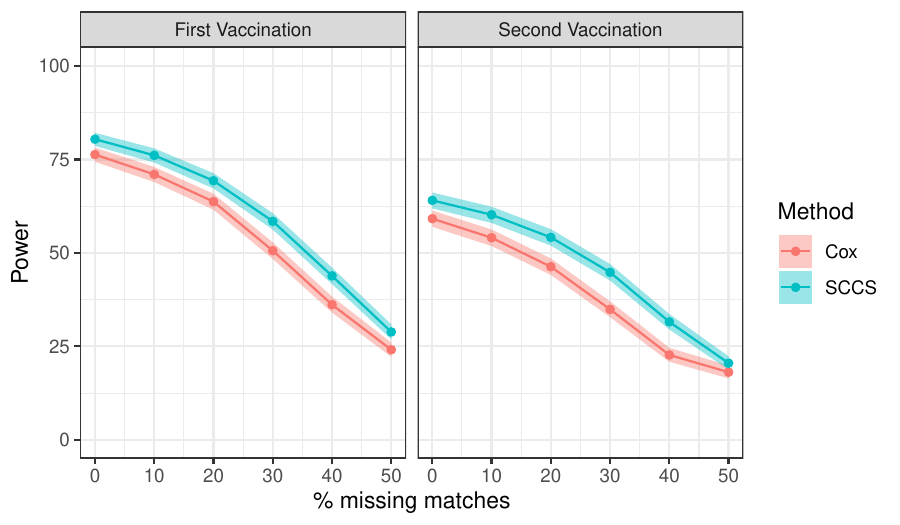}
	\caption{Power of the two evaluation methods for detecting the effect of mRNA vaccination on the occurrence of myocarditis for different proportions of missing matches ($\alpha = 0.05$) with 95\% exact binomial confidence intervals.}
	\label{fig::myokarditis_power}
\end{figure}

The methods are more similar in terms of statistical power, as depicted in figure~\ref{fig::myokarditis_power}. With no missing matches the power is approximately 80\% to detect a statistically significant effect for the first vaccination for the SCCS method and slightly lower for the Cox model. With an increasing percentage of missing matches the power is going down proportionally for both methods. The decrease in power is gradual at first, with 10\% missing matches having only a small impact and grows steeper with increasing percentages. At 50\% missing matches, the power to detect a statistically significant effect for the first vaccination has decreased to approximately 25\% for both methods.
\par\medskip

\begin{figure}[!htb]
	\centering
	\includegraphics[width=1\linewidth]{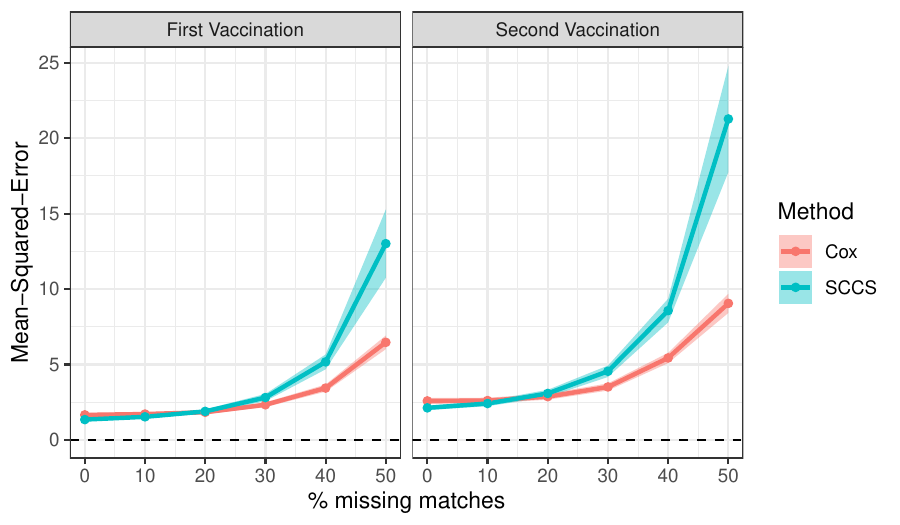}
	\caption{Mean-squared error of the two evaluation methods in detecting the effect of mRNA vaccination on the occurrence of myocarditis for different proportions of missing matches with 95\% confidence intervals.}
	\label{fig::myokarditis_mse}
\end{figure}

The MSE as depicted in figure~\ref{fig::myokarditis_mse}, shows a slightly different picture of the performance of the two methods. With no missing matches, the mean squared error is very similar in both methods with the SCCS method showing a slightly smaller MSE. As the percentage of missing matches increases, the MSE grows bigger for both methods. Interestingly, the MSE increases more sharply for the SCCS method. These results hold for both the first and the second vaccination.

\FloatBarrier

\section{Discussion}

In this article we studied the impact of random record-linkage errors on the bias and statistical power in Covid-19 vaccine safety analysis. We compared the SCCS method and a Cox model with the vaccination status modelled as a time-varying covariate using a Monte-Carlo simulation study. The simulations showed that without record-linkage errors, both methods produce unbiased results with similar statistical power. With a growing percentage of missing matches however, the performance of the Cox model deteriorates, showing a significant amount of bias and loss of statistical power. The SCCS method on the other hand, produces unbiased results even with a large proportion of missing matches, albeit with a similar loss of statistical power.
\par\medskip
The reason for this rather profound difference in performance of these two methods lies in their design. When using the Cox model, unvaccinated individuals are compared with vaccinated individuals to estimate the hazard-ratio of interest. When matches are missing due to errors in the record-linkage process, these missing matches are included in the Cox model as unvaccinated individuals, which results in an underestimation of the true vaccination effect. The SCCS method does not have this problem, because it compares the time at risk after the vaccination with the rest of the persons' observed time for each individual \parencite{Farrington1995}. Consequently, unvaccinated people are excluded from the estimation since they have no time at risk to perform a comparison with. Therefore, random missing matches only result in a smaller sample size for the SCCS method, which decreases the statistical power, but has no effect on bias if the sample does not get unreasonably small. Similar results have been shown in other simulation studies performed by \textcite{Glanz2006} and \textcite{Takeuchi2018}, although these authors were not concerned with record-linkage errors directly.
\par\medskip
The key assumption for this result to hold is that the missing matches occur randomly, meaning that the probability for a missing match is independent of any characteristics of the individuals included in the databases to be linked. This assumption may not hold in practice, as shown in different studies \parencite{Boyd2017, Miller2017, Randall2018}. For example, some groups of people may have names that are more difficult to spell than others, leading to a higher probability of missing matches \parencite{Schnell2023}. If these people also share some characteristics such as age, gender or migration background, that are associated with the outcome of interest, selection bias may be introduced into the analysis when simply ignoring the missing matches \parencite{Parrish2017}. Future studies could introduce fixed and/or time-varying covariates associated with both the probability of a missing match and the probability of developing the disease of interest into our simulation model to further study the effects of non-random record-linkage errors.
\par\medskip
Despite this limitation, our results have multiple implications. The first one is the general result that when performing vaccine safety analysis using data that was created using pseudo-identifier based record-linkage methods, it may be preferable to use case-only designs such as the SCCS method as opposed to a Cox model. The performance of SCCS is comparable to the Cox model in the absence of record-linkage errors and significantly better in the presence of random missing matches. This result also extends to more general situations where information about risk periods has not been recorded for some individuals. For example, missing exposure data for a subset of individuals in an registry study is functionally equivalent to the type of missing matches presented here. It is important to note, however, that this result only holds if all time periods under risk of a person are unknown. If only some are not recorded, the relative risk would be underestimated because the person would still be included in the analysis, but some person-time at-risk would be treated as time not at-risk.
\par\medskip
Additionally, our results have direct implications for the analysis of Covid-19 vaccine safety using German health-care data. Assuming 10 to 20\% random missing matches, the bias is negligible when using the SCCS method. The loss of statistical power shown in this study could be outweighed by the much larger sample size of the real dataset that could be created by linking German health-care data with a database on vaccination information. Due to the large sample size, a well-performed record-linkage study in Germany could identify even very rare Covid-19 vaccine side-effects that could not be detected using the smaller, but already linked datasets of other countries such as England \parencite{HippisleyCox2021, Patone2022} or Israel \parencite{Barda2021}.

\section*{Acknowledgements}

We want to thank Dr. Hans H. Diebner for his extensive input and scientific discussion.

\FloatBarrier
\newpage

\printbibliography

\appendix

\newpage

\section{List of Assumptions made for the Data Generation}

It was assumed that,

\begin{itemize}
	\item the modelled process is appropriately represented via discrete time.
	\item the baseline hazard is constant over time.
	\item the data of the Robert Koch Institute concerning vaccinations is correct.
	\item the proportion of vaccine types employed is constant over time.
	\item the effect of the vaccination on adverse events follows a step-function.
	\item the occurrence of events is independent of prior events (except when using the ``immunity''-parameter).
	\item no confounders in relation to the causal effect of the vaccination on adverse events exist.
	\item every person only receives one type of vaccine.
	\item the time period between administering vaccinations always matches the official recommendations precisely.
	\item that every person has the same probability of getting vaccinated (depending on time).
\end{itemize}

\newpage

\section{Additional Information about Simulation Parameters}

\begin{figure}[!htb]
	\centering
	\includegraphics[width=\linewidth]{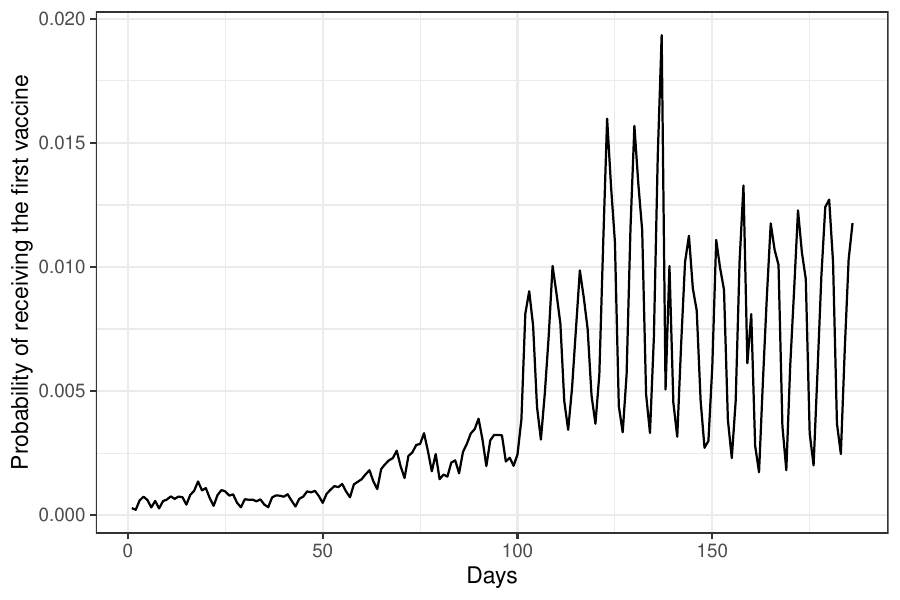}
	\caption{Probability of receiving the first dose of any Covid-19 vaccine as a function of time since the first vaccine was administered as used in the simulation study. Used as reported by the Robert Koch Institute \parencite{RKI2023} as of the first of July 2021.}
\end{figure}

\begin{table}[!htb]
	\centering
	\label{tab::vacc_types}
	\begin{tabular}{ll}
		\toprule
		Vaccine Type & Probability \\
		\midrule
		Pfizer BioNTech & 0.6777 \\
		Moderna & 0.08083 \\
		AstraZeneca & 0.1993 \\
		Janssen & 0.04216 \\
		\bottomrule
	\end{tabular}
	\caption{The probability distribution utilized to determine which type of vaccine a person received in the simulation study. Used as reported by the Robert Koch Institute \parencite{RKI2023} as of the first of July 2021.}
\end{table}

\FloatBarrier
\newpage

\section{Additional Results}

\begin{table}[!htb]
	\centering
	\begin{tabular}{lccrrrr}
		\toprule
		Model & Vaccination & \% Missing Matches & $\widehat{Bias}$ & $SE(\widehat{Bias})$ & $\widehat{MSE}$ & $SE(\widehat{MSE})$ \\ 
		\midrule
		Cox & First & 0 & 0.018 & 0.029 & 1.653 & 0.061 \\ 
		Cox & First & 10 & -0.070 & 0.029 & 1.710 & 0.060 \\ 
		Cox & First & 20 & -0.201 & 0.030 & 1.827 & 0.066 \\ 
		Cox & First & 30 & -0.333 & 0.033 & 2.325 & 0.081 \\ 
		Cox & First & 40 & -0.476 & 0.040 & 3.432 & 0.119 \\ 
		Cox & First & 50 & -0.579 & 0.055 & 6.464 & 0.230 \\ 
		Cox & Second & 0 & 0.046 & 0.036 & 2.581 & 0.101 \\ 
		Cox & Second & 10 & -0.070 & 0.036 & 2.604 & 0.095 \\ 
		Cox & Second & 20 & -0.255 & 0.037 & 2.862 & 0.101 \\ 
		Cox & Second & 30 & -0.455 & 0.041 & 3.504 & 0.122 \\ 
		Cox & Second & 40 & -0.641 & 0.050 & 5.429 & 0.174 \\ 
		Cox & Second & 50 & -0.819 & 0.065 & 9.049 & 0.324 \\ 
		SCCS & First & 0 & -0.016 & 0.026 & 1.342 & 0.047 \\ 
		SCCS & First & 10 & -0.005 & 0.028 & 1.525 & 0.054 \\ 
		SCCS & First & 20 & 0.017 & 0.031 & 1.885 & 0.068 \\ 
		SCCS & First & 30 & 0.076 & 0.037 & 2.804 & 0.122 \\ 
		SCCS & First & 40 & 0.132 & 0.051 & 5.168 & 0.252 \\ 
		SCCS & First & 50 & 0.261 & 0.080 & 13.009 & 1.156 \\ 
		SCCS & Second & 0 & 0.010 & 0.033 & 2.120 & 0.075 \\ 
		SCCS & Second & 10 & 0.017 & 0.035 & 2.408 & 0.090 \\ 
		SCCS & Second & 20 & 0.030 & 0.039 & 3.076 & 0.124 \\ 
		SCCS & Second & 30 & 0.070 & 0.048 & 4.547 & 0.192 \\ 
		SCCS & Second & 40 & 0.135 & 0.065 & 8.573 & 0.401 \\ 
		SCCS & Second & 50 & 0.335 & 0.103 & 21.277 & 1.792 \\ 
		\bottomrule
	\end{tabular}
	\caption{Performance Criteria and associated Monte-Carlo simulation standard errors for both models, vaccinations and all considered proportions of missing matches. All values are rounded to the third digit. Estimated using 2000 simulation repetitions.}
\end{table}

\begin{figure}[!htb]
	\centering
	\includegraphics[width=\linewidth]{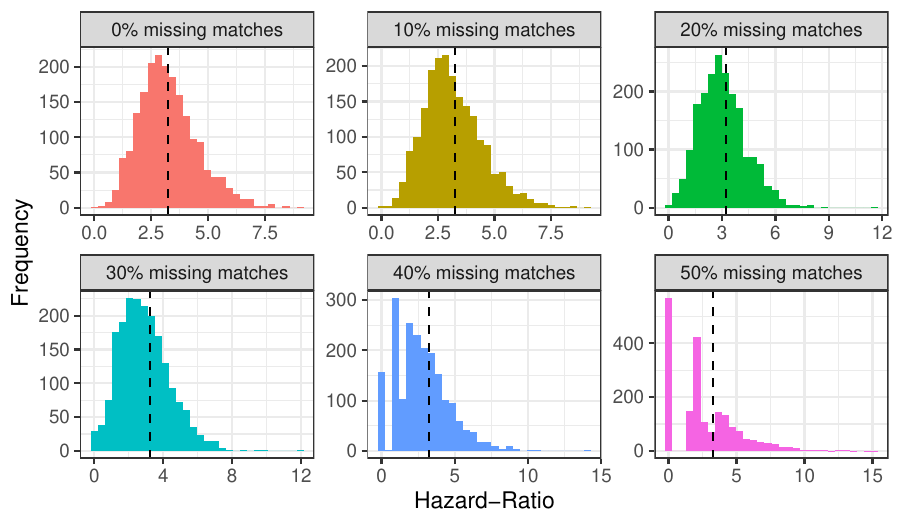}
	\caption{Histograms of the Hazard-Ratios for the first mRNA vaccination estimated using a Cox model over all simulation replications.}
\end{figure}

\begin{figure}[!htb]
	\centering
	\includegraphics[width=\linewidth]{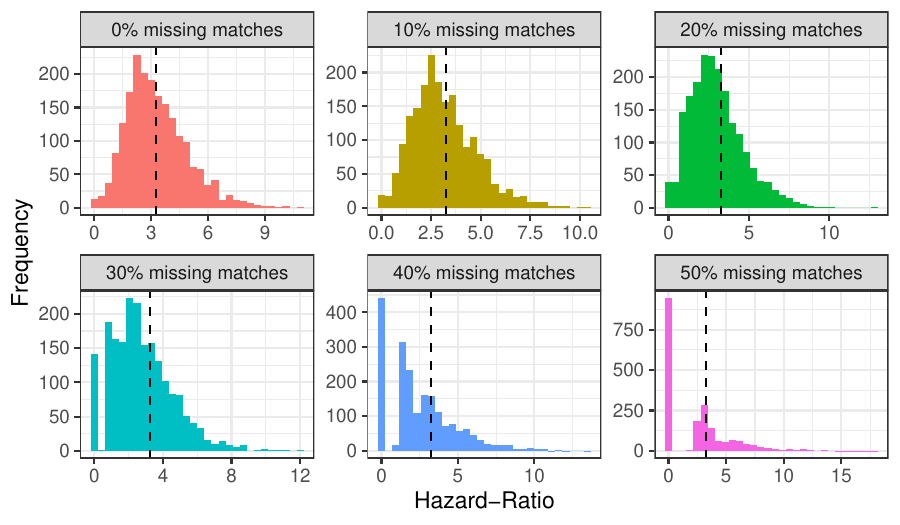}
	\caption{Histograms of the Hazard-Ratios for the second mRNA vaccination estimated using a Cox model over all simulation replications.}
\end{figure}

\begin{figure}[!htb]
	\centering
	\includegraphics[width=\linewidth]{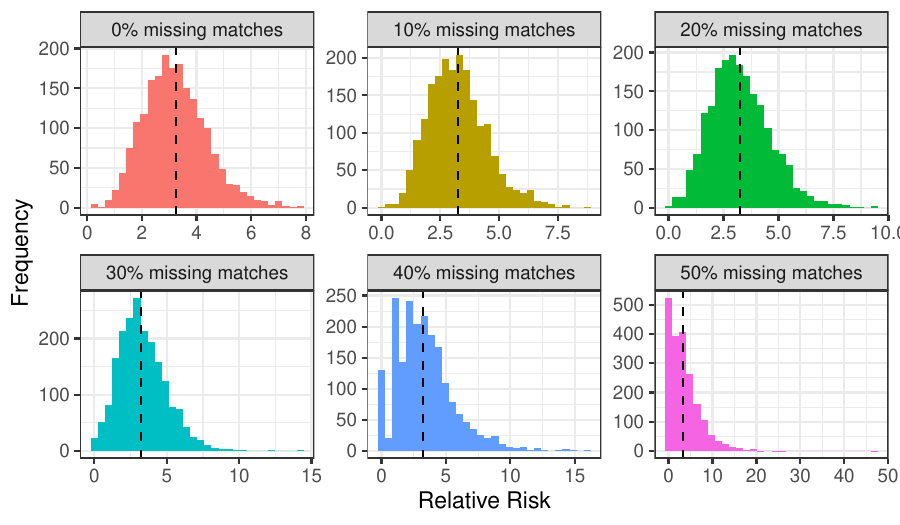}
	\caption{Histograms of the relative risks for the first mRNA vaccination estimated using the self-controlled case serie method over all simulation replications.}
\end{figure}

\begin{figure}[!htb]
	\centering
	\includegraphics[width=\linewidth]{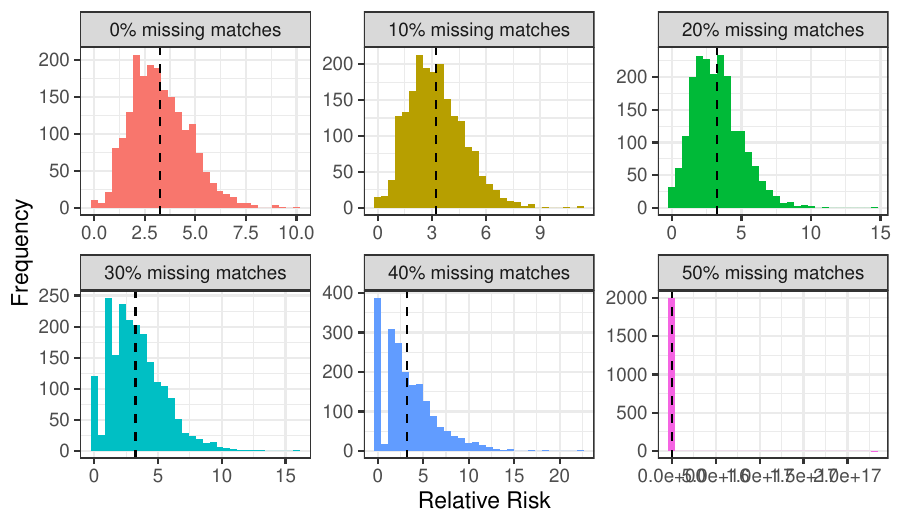}
	\caption{Histograms of the relative risks for the second mRNA vaccination estimated using the self-controlled case serie method over all simulation replications.}
\end{figure}

\end{document}